\documentclass[prc,showpacs,twocolumn,floatfix]{revtex4}
\usepackage{graphicx}
\usepackage{dcolumn}
\usepackage{bm}
\usepackage{multirow}
\usepackage{color}
\usepackage{amssymb}

\begin{document}

\title{Role of higher order couplings in the presence of kaons in relativistic
mean field description of neutron stars}
\author{Neha Gupta}
\email{nehaguptaiitr@gmail.com}
\author{P. Arumugam}

\affiliation{Department of Physics, Indian Institute of Technology Roorkee, Roorkee - 247 667, India}

\date{\today}

\begin{abstract}
We discuss the role of higher order couplings in conjunction with kaon condensation
using  recent versions of relativistic mean field models. We focus on an interaction (G2)  in which all parameters are obtained by fitting finite
nuclear data and successfully applied to reproduce variety of nuclear properties. Our results show that the higher order couplings play a significant role  at higher densities where kaons dominate the behavior of equation of
state.  We compare our results with other interactions (NLl, NL3, G1 and FSUGold) and show that the new couplings bring down the mass of neutron
star (NS), which is further reduced in the presence of kaons to yield results
consistent with present observational constraints.  We show that the composition of NS vary with the parameter sets.   
\end{abstract}

\pacs{
26.60.-c, 
26.60.Kp, 
13.75.Jz, 
97.60.Jd  
}
\maketitle

\section{Introduction \label{section:intro}} 
Neutron stars (NS) provide us with opportunities to  probe the properties of matter at extremely high densities, and have proven to be fantastic test
bodies for theories involving general relativity. In a broader perspective NS provide access to the phase diagram of matter at extreme densities and temperatures, which is the basic for understanding  very early Universe and several other astro-physical phenomena. The observational quantities of primary astrophysical interest are the maximum mass and the typical radius of a NS.
Neuron stars are detected as pulsar+NS or pulsar+white-dwarf or X-ray binaries.  Recent observations of pulsars and X-ray binaries suggest that the   maximum mass of a NS lies between $1-2\ M_\odot$ \cite{Steiner33,expt_m,expt_m1,expt_m2,expt_m3,demorest},
where $M_\odot$ is the solar mass. 

To understand the observables of NS, various theoretical  models have been
developed and can be grouped into three broad categories \cite{lattimer:426}: nonrelativistic potential models \cite{nonrel,nonrel1,nonrel2}, relativistic field theoretical models \cite{rel} and Dirac-Brueckner-Hartree-Fock models
\cite{dbhf,dbhf1}. In each of these models, the addition of hyperon \cite{rel_hyperon}
or kaon or pion  \cite{nonrel_pion,rel_hy_quark,glen_kaon,glen_prl}
 or quarks \cite{rel_hy_quark} or their combinations, will soften the equation of state (EoS) and hence lower the maximum  mass of NS.
In all of these models, coupling constants and
unknown meson masses are treated as effective parameters adjusted to fit the empirical
quantities (saturation density $\rho_0$, binding energy $E/A$, compression
modulus $K_\infty$, effective nucleon mass $m^*_n$ and asymmetry energy $J$) at
nuclear saturation. The compression modulus ($K_\infty)$ defines the curvature of the EoS\ at
saturation density and its value will be reflected in the high density behaviour
(stiffness or softness)
of the EoS.  Thus $K_\infty$ will have a direct bearing on the maximum
mass. Interestingly, the study of isoscalar giant resonances can reveal rich
information about nuclear compressibility \cite{Uchida:12,Li:162503}.  The EoS and hence the NS
radius can directly be linked to the neutron skin thickness in heavy nuclei
\cite{horowitz:062802,FSU_para,sharma:23}.  These strong links between the NS structure and finite nuclear properties
prove that the knowledge in these two areas complement each other.
  
 One among the very-well tested model in finite nuclei is the nonlinear $\sigma-\omega-\rho$
 model, widely mentioned as  relativistic mean field (RMF) model \cite{Gambhir:132}. The RMF model constructed with idea of renormalizability, produces accurate results for spherical and
deformed nuclei but it gives a very stiff EoS for infinite nuclear matter. Recently, inspired by effective field theory (EFT) Furnstahl, Serot and Tang \cite{furnstahl} abandoned the idea of renormalizability and extended the RMF theory, with the systematic inclusion of new interactions by adding new
terms to the model Lagrangian. This effective field theory
motivated RMF (E-RMF) model calculations  explained finite nuclei and nuclear matter with exactly same parameters and with significant accuracy in both the
cases \cite{ermf}. This approach can be considered as a salient step towards a unified and accurate theory for finite nuclei as well as for infinite nuclear matter. Extension of this
model with the inclusion of the kaon ($K^-$) is the central interest of present
work.
 
In NS, $K^-$-condensation is one of the several possible transitions that could exist
at high density. As the  density of NS increases,  the chemical potential of the negative electric charge ($\mu_e$) also increases with the same rate
of proton number density. Simultaneously, the effective mass of an in-medium
$K^-$ will decrease, due to the attractive interaction between $K^-$
and nuclear matter. Therefore at a particular density (when $\mu_e$ is greater
than energy of $K^-$), $K^-$ replaces electron as neutralizing agent in charge-neutral matter.
After the demonstration of interaction of $K^-$ with the nuclear medium by
Kaplan and Nelson \cite{kap}, this topic acquired enormous interest. Glendenning and Schaffner-Bielich \cite{glen_prl,glen_kaon}, explained the
interaction of $K^-$ by coupling them to meson fields using a minimal coupling.
This approach was followed by several other theoretical groups using RMF models \cite{sanjay_kaon,sanjay_kaon_t,kaon_wang} and has been adopted in
this work as well.   

In the present work we investigate the condensation from non-kaonic to kaonic
phase, with different Lagrangians. In the presence of such a transition,
we study the effect of higher order couplings on the EoS and hence on the NS properties. This work differs from many previous works in the sense that
all the parameters used in E-RMF model were fitted to reproduce the observables
of finite nuclei. Also, the E-RMF model is well tested throughout the nuclear chart by explaining several nuclear properties \cite{DelEstal:443,DelEstal:024314,DelEstal:044321,Sil:044315,
Roca-Maza:044332,Shukla:034601}. We show in this paper
that this model explains the recent observations of NS as well.

In the section following this introduction, we describe the Lagrangian and field equations,
both for non-kaonic and kaonic phases. This is followed by the expressions
for energy density and pressure, which define the EoS. In section \ref{section:parameters} we discuss the parameters used in our calculation. Our results and discussions are presented in section \ref{section:results} which is followed by the summary along with the conclusions drawn from present work. 

\section{Effective field theory motivated relativistic mean field model with kaons \label{section:theory}}
In this section we briefly sketch the E-RMF model
by presenting the model Lagrangian   \cite{ermf,furnstahl}. We then show how the physical quantities, that will
determine the  composition of the NS, can be obtained self consistently.

The effective Lagrangian, obtained by curtailing terms irrelevant to nuclear
matter in the E-RMF Lagrangian, can be written as
\begin{eqnarray}
\mathcal{L} &=& \bar\psi[ g_\sigma\sigma-\gamma^\mu (g_\rho R_{\mu}+g_\omega V_\mu)]\psi \nonumber\\&&+\frac{1}{2}\left(1+\eta_{1}\frac{g_\sigma\sigma}{m_n}+\frac{\eta_2}{2}\frac{g_{\sigma}^{2}\sigma^2}{m_n^2}\right)m_{\omega}^{2}V_\mu
 V^{\mu}\nonumber\\&&+\frac{1}{4!}\zeta_0g_{\omega}^{2}(V_\mu
 V^\mu)^2+\left(1+\eta_\rho\frac{g_\sigma \sigma}{m_n}\right)m_{\rho}^{2}\text{tr}( R_\mu R^{\mu})\nonumber\\&&-m_{\sigma}^{2}\sigma^2\left(\frac{1}{2}+\frac{\kappa_3g_\sigma\sigma}{3!m_n}+\frac{\kappa_4g_{\sigma}^{2}\sigma^2}{4!m_n^2}\right),
\end{eqnarray}
where the scalar, vector and isovector meson fields and the nucleon field  are denoted  by $\sigma$, $V_\mu$, $R_\mu$ and $\psi$ respectively. $m_\sigma,\
 m_\omega,$ and $m_\rho$ are the corresponding meson masses and $m_n(= m_p)$ is the nucleon mass.  The symbols $g_\sigma$, $g_\omega$, $g_\rho$, $\kappa_3$, $\kappa_4$, $\eta_1$, $\eta_2$,
$\eta_\rho$ and $\zeta_0$ denote the various coupling constants.
More
details of the Lagrangian are explained explicitly in Ref.~\cite{furnstahl}. 
  
Now we extend the E-RMF model by including a kaon ($K^-$)-nucleon interaction term.
For simplicity, we choose that the kaon is coupled to the meson field with minimal coupling \cite{glen_kaon}.
In this way, interactions of mesons with nucleons and kaons are treated in the same footing.

The Lagrangian for the kaon part reads
\begin{equation}\label{kaon}\mathcal{L}_K=D_\mu^*K^*D^\mu K-m_K^{*2}K^*K,
\end{equation}
where the vector fields are coupled to kaons via
the relation\begin{equation}D_\mu=\partial_\mu+ig_{\omega K}V_\mu+ig_{\rho K}\tau_3\cdot R_\mu,
\end{equation}
and $m^*_K$ is the effective mass of kaon.
The scalar field is coupled to kaons in a way  analogous to the minimal coupling scheme
\cite{glen_kaon} of the vector fields:
\begin{equation}m_K^*=m_K-g_{\sigma K}\sigma,\\ 
\end{equation}
where $m_K=495$ MeV.
Note that in the mean field approximation, only the time components of
the vector fields $V_0$ contribute and charge conservation implies that only the third component in isospin-space of the isovector meson field $R_0$ does not vanish.

The dispersion relation for $s$-wave condensation $(\vec k=0)$, for $K^-$ is 
\begin{equation}\label{eq:ke}\omega_K=m_K-g_{\sigma K}\sigma-g_{\omega K}V_0-g_{\rho K}R_0,
\end{equation}
with $K^-$ mesons having isospin projection ${-1}/{2}$.
$\omega_K$ represents the $K^-$ energy and is linear in the meson field.
 One can note that with the increase in density, $\omega_K$ decreases.

In the presence of $K^-$, equations of motion for the meson fields are 
\begin{eqnarray}
m_{\sigma}^{2}\sigma&=&g_\sigma\rho_s-\frac{m^{2}_{\sigma}g_\sigma\sigma^2}{m_n}\left(\frac{\kappa_3}{2}+\kappa_4\frac{g_{\sigma}\sigma}{3!m_n}\right)
+\eta_\rho\frac{g_\sigma}{2m_n}m_{\rho}^2{R_0}^{2}\nonumber\\&&+\frac{1}{2}\left(\eta_1+\eta_2\frac{g_{\sigma}\sigma}{m_n}\right)\frac{g_\sigma}{m_n}m_{\omega}^{2}V^{2}_0+g_{\sigma
K}\rho_K,\nonumber
\end{eqnarray}
\begin{eqnarray}
m_{\omega}^{2}V_0&=&g_{\omega}(\rho_p+\rho_n)-\left(\eta_1+\frac{\eta_2 g_{\sigma}\sigma}{2m_n}\right)\frac{g_\sigma\sigma}{m_n}
m_{\omega}^{2}V_0\nonumber\\&&-\frac{1}{3!}\zeta_0g_{\omega}^2V_0^3-g_{\omega K}\rho_K,\nonumber
\end{eqnarray}
\begin{equation}\label{eq:fields}
m_{\rho}^{2}R_0=\frac{1}{2}g_\rho(\rho_p-\rho_n)-\eta_\rho\frac{g_\sigma\sigma}{m_n}m_{\rho}^{2}R_0-g_{\rho
K}\rho_K,
\end{equation}
where $\rho_s$ is the scalar density given by
\begin{eqnarray}\rho_s&=&\frac{\gamma}{(2\pi)^3}\sum_{i=n,p}\int_{0}^{k_{fi}}d^3k\frac{m_n^*}{(k^2+m_n^{*2})^{1/2}}
\;,\end{eqnarray}and  $\gamma$ is the spin-isospin degeneracy factor and is equal to 2 (for spin up and spin down).
The nucleon effective mass is defined as in the standard Walecka model 
\begin{equation}m_n^*=m_n-g_\sigma\sigma,
\end{equation}
and the $K^-$ density is given by 
 \begin{equation}
 \rho_K=2m_K^*K^*K=2(\omega_K+g_{\omega K}V_0+g_{\rho K}R_0)K^* K.
 \end{equation}
The proton and neutron chemical potentials are  
\begin{eqnarray}
&& \mu_p=\sqrt{k_{fp}^2+m_n^{*2}}+g_\omega V_0+\frac{1}{2}g_\rho R_0, \text{and}\nonumber\\
&& \mu_n=\sqrt{k_{fn}^2+m_n^{*2}}+g_\omega V_0-\frac{1}{2}g_\rho R_0. \label{eq:cp_np}
\end{eqnarray}

For a NS, in the absence of neutrino trapping, the conservation of baryon and electron chemical potentials leads to \cite{gle}
\begin{eqnarray}
 \mu_n&=&\mu_p+\mu_e,\nonumber\\
 \mu_e&=&\mu_\mu,
 \label{eq:beta_eq}
\end{eqnarray} and
\begin{equation}
q=\rho_p-\rho_e-\rho_\mu-\rho_K,
\label{eq:cd}
\end{equation}
where the first two constraints ensure the chemical equilibrium and the last
one specifies the total charge which vanishes while imposing the charge neutrality.
The total baryon density is
\begin{equation}
\rho=\rho_p+\rho_n
\label{eq:den}
\end{equation}

\begin{table*}
\caption{Parameters  and the saturation properties
for NL1\cite{nl1_para}, NL3\cite{nl3_para},
G1, G2\cite{DelEstal:024314}, and FSUGold\cite{FSU_para}. The parameters $g_\sigma$, $g_\omega$, $g_\rho$, $\kappa_3$ and $\kappa_4$ are calculated from the given
saturation properties using relations suggested in Ref. \cite{gle} with the
exception of FSUGold where all the coupling constants are taken from Ref.~\cite{FSU_para}.}
\label{nucleon}
\centering
\begin{tabular}{c d d d d d}
\hline\hline
& \text{NL1} & \text{NL3} & \text{G1} & \text{G2} & \text{FSUGold}\\[0.5ex]\hline
$m_n$ (MeV) &938 & 939 &939 & 939 & 939\\
$m_\sigma$ (MeV) & 492.25 & 508.194 & 507.06 & 520.206 & 491.5\\
$m_\omega$ (MeV) & 795.36 & 782.501 & 782 & 782 & 783\\
$m_\rho$ (MeV) & 763 & 763 & 770 & 770 & 763\\
$g_\sigma$ & 10.0730 & 10.1756 & 9.8749 & 10.5088 & 10.5924\\
$g_\omega$ &  13.1917 & 12.7885 & 12.1270 & 12.7864 &14.3020\\
$g_\rho$ & 9.8553& 8.9849&8.7886&9.5108&11.7673\\
$\kappa_3$ & 1.8324&1.4841&2.2075&3.2376&0.6194\\
$\kappa_4$ & -7.7099&-5.6596&-10.0808&0.6939&9.7466\\
$\eta_1$&0&0&0.071&0.65&0\\
$\eta_2$&0&0&-0.962&0.11&0\\
$\eta_\rho$&0&0&-0.272&0.390&0\\
$\zeta_0$&0&0&3.5249&2.642&0.06\\
$\Lambda$&0&0&0&0&0.03\\
\hline\\
$\rho_0$ (fm$^{-3}$)&0.154&0.148&0.153&0.153&0.148\\
(E/A)(MeV)&-16.43&-16.299&-16.14&-16.07&-16.3\\
$K_\infty$(MeV)&212&271.76&215&215&230\\
$J$(MeV)&43.6&37.4&38.5&36.4&32.59\\
$m_n^*/m_n$&0.571&0.6&0.634&0.664&0.609\footnotemark[1]\\[0.5ex]
\hline\hline
\end{tabular}
\footnotetext[1]{$m^*_n$ for FSUGold was calculated from the coupling constants.}
\end{table*}

The energy density can be written as 
\begin{equation}\label{eq:ed}\epsilon=\epsilon_N+\epsilon_K,
\end{equation}
where $\epsilon_N$ is the energy density due to nucleons given by
\begin{eqnarray}\label{eq:ed_N}
\epsilon_N&=&\sum_{i=n,p,l}\frac{\gamma}{(2\pi)^3}\int_{0}^{k_{fi}}d^3k\sqrt{ k^2+m_i^{*2}}-\frac{1}{4!}{\zeta_0}{g_{\omega}^2}V_{0}^{4}\nonumber\\&&-\frac{1}{2}\left(1+\eta_1\frac{g_\sigma\sigma}{m_n}+\frac{\eta_2}{2}\frac{g_{\sigma}^2\sigma^{2}}{m_n^2}\right)m_{\omega}^{2}V^{2}_0+g_\omega
V_0(\rho_p+\rho_n)\nonumber\\&&
-\frac{1}{2}\left(1+\eta_\rho\frac{g_\sigma\sigma}{m_n}\right)m_{\rho}^{2}R_{0}^2+\frac{1}{2}g_\rho
R_0(\rho_p-\rho_n)\nonumber\\&&+{m_{\sigma}^{2}}{\sigma^{2}}\left(\frac{1}{2}+\frac{\kappa_3}{3!}\frac{g_\sigma\sigma}{m_n}+\frac{\kappa_4}{4!}\frac{g_{\sigma}^2\sigma^{2}}{m_n^2}\right),
\end{eqnarray}
and the energy density contributed by kaons is
\begin{equation}\label{eq:ed_K}\epsilon_K=2m_K^{*2}K^*K=m_K^*\rho_K.
\end{equation}
 Unlike the energy density, pressure is not directly affected by the inclusion of
$K^-$, but the inclusion of $K^-$ affects the fields and hence the pressure
which reads
\begin{eqnarray}\label{eq:pre}
p&=&\sum_{i=n,p,l}\frac{\gamma}{3(2\pi)^3}\int_{0}^{k_{fi}}d^3k\frac{k^2}{\sqrt{k^2+m_i^{*2}}}+\frac{1}{4!}{\zeta_0}{g_{\omega}^2}V_{0}^{4}\nonumber\\&&+\frac{1}{2}\left(1+\eta_1\frac{g_\sigma\sigma}{m_n}+\frac{\eta_2}{2}\frac{g_{\sigma}^2\sigma^{2}}{m_n^2}\right)m_{\omega}^{2}V_{0}^2\nonumber\\&&
-{m_{\sigma}^{2}}{\sigma^{2}}\left(\frac{1}{2}+\frac{\kappa_3}{3!}\frac{g_\sigma\sigma}{m_n}+\frac{\kappa_4}{4!}\frac{g_{\sigma}^2\sigma^{2}}{m_n^2}\right)
\nonumber\\&&+\frac{1}{2}\left(1+\eta_\rho\frac{g_\sigma\sigma}{m_n}\right)m_{\rho}^{2}R_{0}^2.
\end{eqnarray}
Here $l$ stands for the leptons $ (e^-,\mu^-)$.

\subsection{Non-kaonic phase ($n$, $p$, $e^-$, $\mu^-$)}
In the non-kaonic phase,  vanishing charge density implies $q\equiv 0$ with $\rho_K=0$. We can calculate $\sigma$, $V_0$, $R_0$, $k_{fp}$, $k_{fn}$, $k_{fe}$, and $k_{f\mu}$ by using Eqs.~(\ref{eq:fields}),
(\ref{eq:beta_eq}), (\ref{eq:cd}) and (\ref{eq:den}), at the chosen baryon density.  When we get the converged solution for the above-listed quantities, the energy density and pressure can be computed from Eqs.~(\ref{eq:ed}) and (\ref{eq:pre}).     

\subsection{Kaonic phase ($n$, $p$, $e^-$, $\mu^-$, $K^-$)}
With the solution of non-kaonic phase in hand, from Eq.~(\ref{eq:ke}) we can calculate kaon energy  which keeps decreasing as we increase density, while $\mu_e$ increases. When the condition $\omega_K= \mu_e $ is first achieved, the kaon will occupy a small fraction of the total volume and the charge density corresponding to kaonic phase, $q\equiv0.$ We can calculate $\sigma$, $V_0$, $R_0$, $k_{fp}$, $k_{fn}$, $k_{fe}$, $k_{f\mu}$ and $\rho_K$ by using Eqs.~(\ref{eq:fields}), (\ref{eq:beta_eq}) (\ref{eq:cd}) and (\ref{eq:den}) with the condition $\omega_K= \mu_e $, for any chosen baryon density. After getting these solutions, we can calculate energy density and pressure for kaonic phase, using Eqs.~(\ref{eq:ed}) and (\ref{eq:pre}). 

\section{ Choice of parameters\label{section:parameters}} In the present work we have chosen five sets
of parameters namely NL1 \cite{nl1_para}, NL3 \cite{nl3_para},
G1, G2 \cite{DelEstal:024314} and FSUGold \cite{FSU_para}. The first two parameter
sets correspond to the standard RMF model with very different compressibilities
[$K_\infty$(NL1) = 212 MeV and $K_\infty$(NL3) = 271.76 MeV]. In the case of FSUGold, there
are two more  coupling parameters $\Lambda $ and $\zeta_0$ in  comparison to NL1 and NL3 parameter sets. These parameters represent the strength of the self-interaction of vector field ($\zeta_0$) and the isoscalar -isovector mixing ($\Lambda$).  The interactions G1 and G2 correspond to the complete E-RMF Lagrangian  discussed
in the earlier section. One can note that in  comparison to NL1 and NL3,  G1 and G2 have four more coupling constants ($\eta_1, \eta_2, \eta_\rho,\
$ and $\zeta_0$). All the coupling constants are obtained by fit to several
properties of finite nuclei \cite{furnstahl}. It is worth
to mention that in this case no parameter is treated as adjustable and the
fitting does not involve any observation at densities above the saturation
value.
Since
the expectation value of the $R_0$ field is typically an order of magnitude smaller than that
of the $V_0$ field, in the E-RMF model, the nonlinear $R_0$ couplings were
retained  only through third order \cite{furnstahl}.  However, it has been
shown that the isoscalar-isovector mixing ($\Lambda (g_\omega
V_0)^2(g_\rho R_0)^2$) is useful to modify the
neutron radius in heavy nuclei while making very small
changes to the proton radius and the binding energy \cite{horowitz:062802,FSU_para}.  

In the effective Lagrangian approach adopted here, knowledge of two distinct
sets of coupling constants --- one parametrizing the nucleon-nucleon interaction
and one parametrizing the kaon-nucleon interactions --- is required for numerical
computations. We discuss each of these in turn.
\subsection{Nucleon coupling constants}
The symbols $g_\sigma$, $g_\omega$, $g_\rho$, $\kappa_3$, $\kappa_4$, $\eta_1$, $\eta_2$,
$\eta_\rho$, $\zeta_0$ and $\Lambda$ denote the nucleon coupling constants.
At times these constants bear different values in literature for the same
parameter set. For example, for the set NL1 different numbers are quoted
in Refs.~\cite{nl1_para} and \cite{nl3_para} but they yield same saturation
properties.  Among the above-mentioned coupling constants,  $g_\sigma$, $g_\omega$, $g_\rho$, $\kappa_3$ and $\kappa_4$ can be written algebraically in terms of empirical quantities:
$\rho_0$, $E/A$, $K_\infty$, $J$, $m_n^*$ and vice versa \cite{gle}.   Using these relations we have calculated
$g_\sigma$, $g_\omega$, $g_\rho$, $\kappa_3$ and $\kappa_4$ in the case of
 NL1 \cite{nl1_para}, NL3 \cite{nl3_para},
G1 and G2 \cite{DelEstal:024314}. These calculated
parameters, and other parameters taken from corresponding references are listed in Table \ref{nucleon}.
In case of FSUGold, all the coupling constants are taken
from Ref.~\cite{FSU_para}.

 \subsection{Kaon coupling constants}
In order to investigate the effect of  kaons on the high density matter, the kaon-nucleon coupling constant has to be specified. The laboratory experiments give information only about kaon-nucleon interaction
in free space. In this work we  mainly focus on the NS (densities $\gg\ \rho_0$), and therefore the kaon-nucleon interaction determined from experiment
need  not be appropriate for our calculations. The  interaction of omega and rho mesons with kaon ($g_{\omega K}$ and $g_{\rho K}$) can be determined using a simple quark and isospin counting argument \cite{glen_kaon} given by,
 \begin{equation}g_{\omega K}=\frac{1}{3}g_\omega \text{ and }
g_{\rho K}=\frac{1}{2}g_\rho \;.
\end{equation}
 We can specify the interaction of sigma meson with kaon ($g_{\sigma K}$) using its relation with the optical potential of a single kaon
in infinite matter ($U_K$):
\begin{equation}U_K=-g_{\sigma
K}\sigma(\rho_0)-g_{\omega K} V_0(\rho_0),
\end{equation}
where typically we have $-80$ MeV $\lesssim U_K\lesssim-180$ MeV \cite{koch,waas}.
\section{Results and Discussions\label{section:results}}
The quality of different interactions in explaining symmetric matter and pure neutron matter properties at higher densities than the normal nuclear densities has been
discussed in Refs.~\cite{ermf,fattoyev.055803}, by comparing the calculated
pressure with the experimental  data \cite{expt_data}.  This comparison   clearly suggests a softer EoS for symmetric matter,
which could be obtained only with higher
order couplings in the Lagrangian.  The pure
neutron matter data, though model dependent, also was found to favour a softer EoS. One important observation
highlighted in Ref. \cite{ermf}, is that the softness in EoS from G1, G2
and FSUGold is mainly due to the new couplings and not due to the difference
in compressibility as usually perceived. For example, the compressibility from NL1 ($K_\infty$=212) and G1, G2 ($K_\infty$=215) are very similar but the EoS
at higher densities are completely different. NL1 ($K_\infty$=212) and NL3
($K_\infty$=271.76) yield different compressibilities but their EoS are more similar. The role of higher order couplings in softening the EoS has been discussed in Ref.~\cite{fattoyev.055803} for the case of FSUGold. EoS
from FSUGold is softer than that of G2 because of the large and positive
$\kappa_4$ value as well as the introduction of isoscalar-isovector coupling ($\Lambda$). In fact it has been clearly demonstrated \cite{horowitz:5647,horowitz:062802} that $\Lambda$ softens
the symmetry energy considerably. 

The success of G1 and G2 in explaining the high density EoS constrained by
phenomenological flow analysis \cite{expt_data}, has been carried forward in explaining NS properties as well
\cite{ermf}. FSUGold also has been successful in explaining both these features
\cite{FSU_para,fattoyev.055803}.  Henceforth we discuss
the role of new  parameters of extended RMF models on the onset and effect of kaon condensation
in NS.  Among the different parameter sets
considered here (Table \ref{nucleon}) G2 and FSUGold have a positive quartic
scalar self-coupling ($\kappa_4$), which is more meaningful than a negative one \cite{ermf,kappa_4}. For this reason we prefer the set G2, when we have to choose between G1 and G2. 

\begin{figure}[t]
\includegraphics[width=.99\columnwidth]{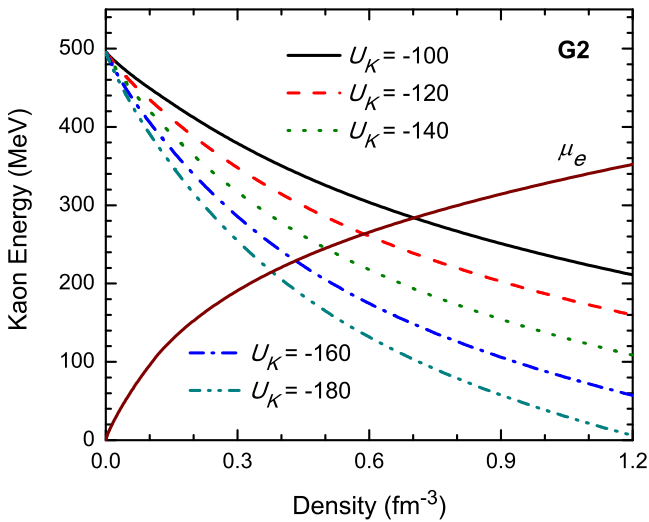}
\caption{(Color online) The density dependence of the $K^-$ energy ($\omega_K$)  in NS matter for different
optical potentials ($U_K$ in MeV) calculated with G2 parameter set.  The
point at which electron chemical potential ($\mu_e$) intersects $\omega_K$
defines the onset of $K^-$-condensation. }\label{kaonen}
\end{figure}

\begin{figure}[t]
\includegraphics[width=.99\columnwidth]{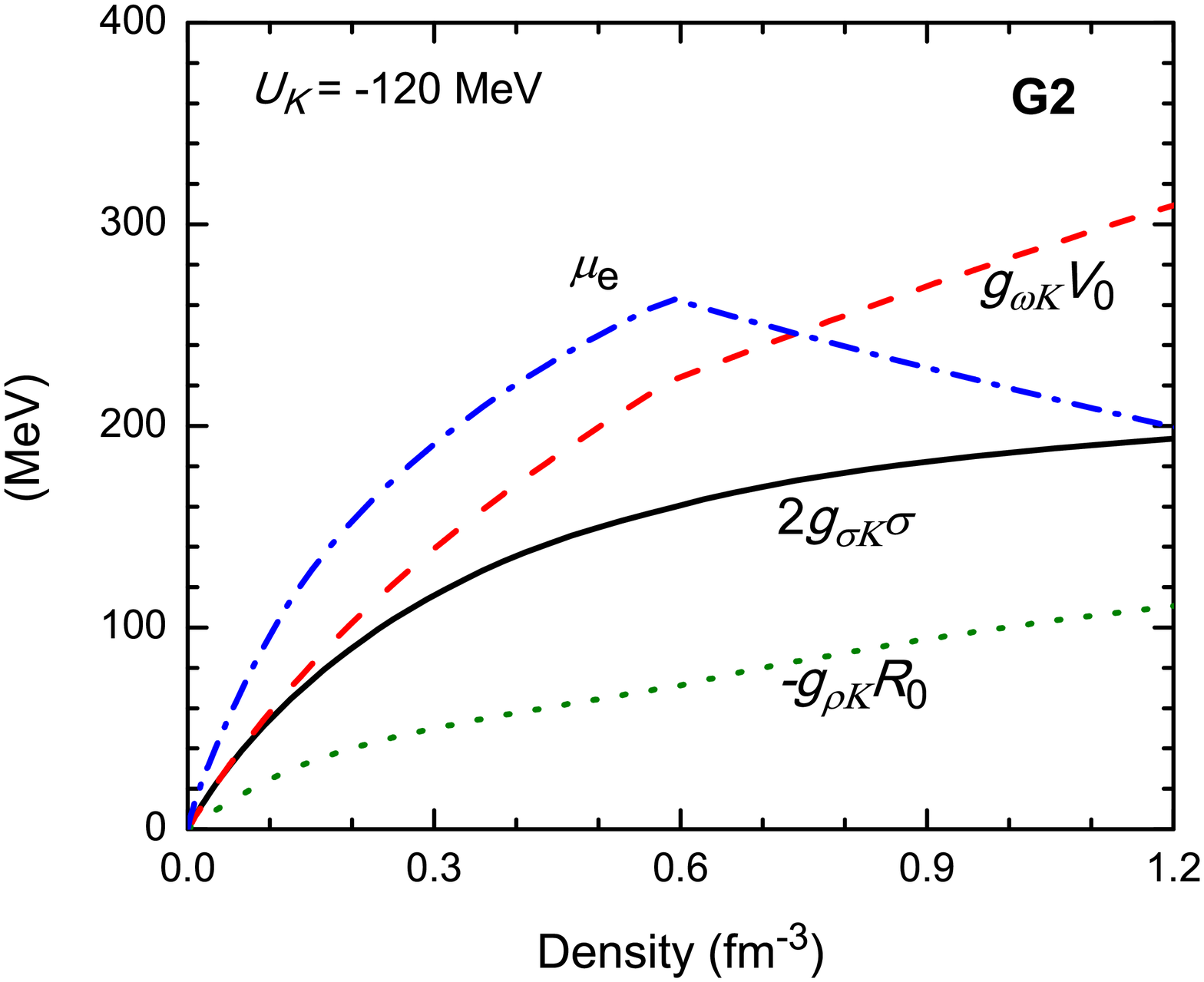}
\caption{(Color online) The density dependence of the scalar $(g_{\sigma K}\sigma)$, vector $(g_{\omega K}V_0)$, and iso-vector $(g_{\rho K} R_0)$ fields
in the NS matter inclusive of kaonic phase, calculated with G2 parameter set. }\label{fieldsG2}
\end{figure}

\begin{figure}[t]
\includegraphics[width=.99\columnwidth]{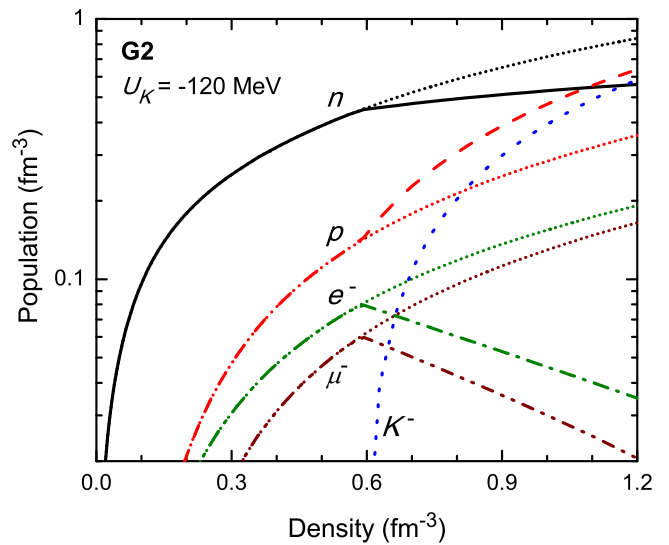}
\caption{(Color online) The relative population of hadrons and leptons in NS as a function of baryon
density calculated with G2 parameter set. The calculations done without considering kaons are represented
by the small-dotted lines.}\label{densityg2}
\end{figure}

\begin{figure}[t]
\includegraphics[width=.99\columnwidth]{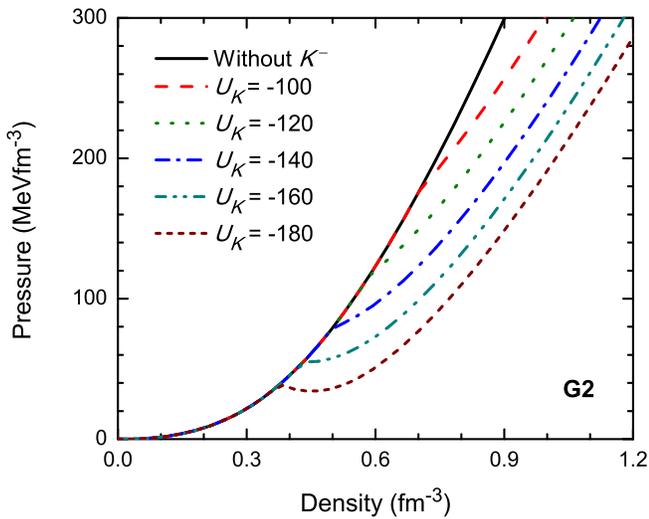}
\caption{(Color online) Pressure in NS matter versus baryon density calculated using different optical potentials and
G2 parameter set.}\label{pressu}
\end{figure}

\subsection{Results from G2}
In Fig.~\ref{kaonen}, we have presented the $K^-$ energy ($\omega_K$) for different optical
potentials, as a function of density along with the variation of electron
chemical potential ($\mu_e$). The $\omega_K$ decreases with increasing density
[Eq.~(\ref{eq:ke})] while $\mu_e$ increases. When  $\omega_K$ is lower than
$\mu_e$,  $K^-$ are favoured (due to attraction between
$K^-$ and nucleon) to replace electrons while
contributing to the  charge neutrality. From Fig.~\ref{kaonen}, it is evident
that the onset of $K^-$-condensation is strongly modified by the strength of
the kaon optical potential ($U_K$). Calculations based on chiral models \cite{waas}
suggest that a value $U_K=-120$ MeV is more appropriate for NS. Hence in most of the further discussions we choose $U_K=-120$ MeV and
look at the other dependencies for the EoS of NS.

The $\sigma$, $\omega$, and $\rho$ fields calculated for NS at $U_K=-120$
MeV are presented in Fig.~\ref{fieldsG2} along with the $\mu_e$. The density at which $K^-$ starts to contribute can be read from
the point where the $\mu_e$ shows a sharp kink ($\sim $ 0.6 fm$^{-3}$). Though
the corresponding kinks in the fields are very subtle, any small change
in the field results in a significant change in energy density and pressure. The presence
of $K^-$ alters the proton-neutron ratio and hence the contribution from the $\rho$ field is enhanced.  This is due to the fact that the processes like $n\rightarrow p+K^-$ are more energetically favoured than processes like $n\rightarrow p+e^-$
at higher densities where $K^-$ energy is decreasing and $\mu_e$ is increasing
[See Fig.~\ref{kaonen}].
From Fig.~\ref{fieldsG2} we can also see that the $K^-$ has very less influence on the $\sigma$ field.  In other words, for the parameters considered  here, the attraction caused by $K^-$ is weak in comparison with that of $\sigma$
mesons. 

The appearance of $K^-$ at higher densities and its role in population (number density) of different particles are presented in Fig.~\ref{densityg2}. As soon as $K^-$-condensation sets in, the population of $K^-$ rapidly increases with density and hence allows
the presence of more protons. Interestingly, at very high densities the number
of protons exceed that of the neutrons, which is not the case if $K^-$ were
not present. This can be seen from the
corresponding deviation from the non-kaonic matter trends. Apart from the
arguments given earlier, the reason for
$p$-$K^-$ pairs being preferred to neutrons is quantitatively very well explained by Glendenning and Schaffner-Bielich \cite{glen_kaon}.
However, the symmetry term in the energy obviously prefer a symmetric matter
and hence hinders the protons to be  populated much more than neutrons.  All the
above-mentioned effects have a significant role to play in determining the EoS
which is discussed in the following text.

Our results for the pressure calculated  with G2 parameter set and with different optical
potentials for $K^-$ are presented in Fig.~\ref{pressu}. On first sight one
can appreciate the strong influence of $U_K$ in softening the EoS, which is similar
to the results of  many previous works. Also for lower values of $U_K$, the graph suggests
 a second order phase transition from a non-kaonic to kaonic phase. Only with a very high value of $U_K(\gtrsim-$160 MeV) one can have a
 first order phase transition. This is rather
 consistent with the observation in Ref. \cite{sanjay_kaon_t} where higher
order interactions are included using the lowest order chiral Lagrangian. However, the role of more general higher order operators was considered to be an
open question which is answered in the present work.  One may expect
 some changes in the onset of $K^-$-condensation if we consider the presence
 of a mixed phase (of kaonic and non-kaonic phases) \cite{glen_kaon}. The presence of such a mixed
phase is normally possible in first order $K^-$-condensation where one gets
a dip in the pressure like the one for $U_K\gtrsim-160$ MeV as shown in Fig.~\ref{pressu}.
This argument implies that the presence of mixed phase is not favoured in our case
for $U_K<-160$ MeV. The main observation from Fig.~\ref{pressu} is that even
in the presence of higher order couplings, the
presence of $K^-$ dramatically softens the EoS and the softness is proportional
to $U_K$.  

\begin{figure}[t]
\includegraphics[width=.99\columnwidth]{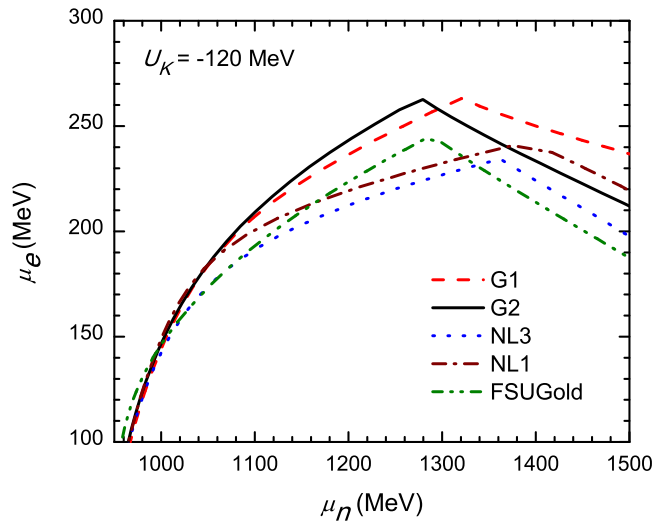}
\caption{(Color online) The electron chemical potential versus the neutron chemical potential
in NS matter calculated with different parameter sets. The kinks represent
transition from non-kaonic to kaonic phase.}\label{cp}
\end{figure}

\begin{figure}[t]
\includegraphics[width=.99\columnwidth]{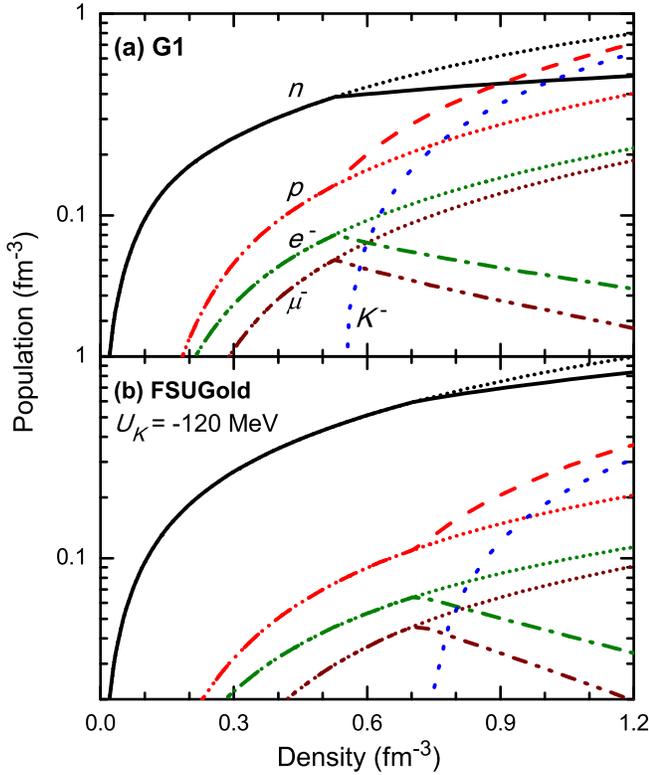}
\caption{(Color online) The relative population of hadrons and leptons in NS matter as function of baryon
density calculated with (a) G1 and (b) FSUGold parameter sets. The calculations done without considering kaons are represented
by the small-dotted lines.}\label{density_g1}
\end{figure}

\begin{figure}[t]
\includegraphics[width=.99\columnwidth]{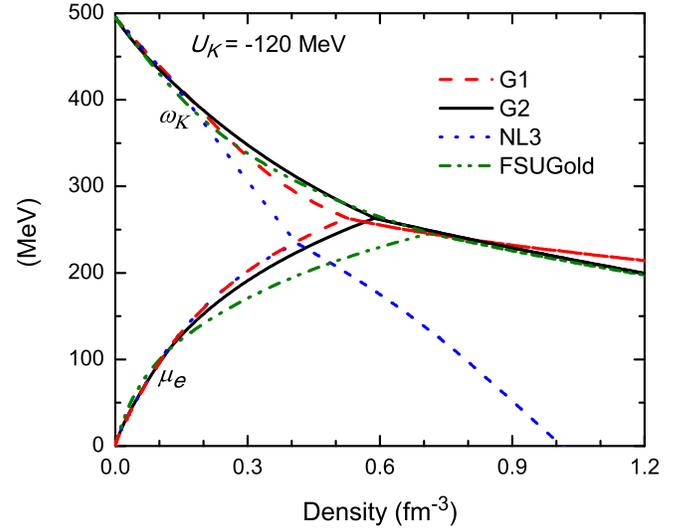}
\caption{(Color online) Density dependence of the electron chemical potential
($\mu_e$) and $K^-$ energy ($\omega_K$) for different parameters. The
point at which $\mu_e$ intersects $\omega_K$
defines the onset of $K^-$-condensation.}\label{kaon_para}
\end{figure}

\begin{figure}[t]
\includegraphics[width=.99\columnwidth]{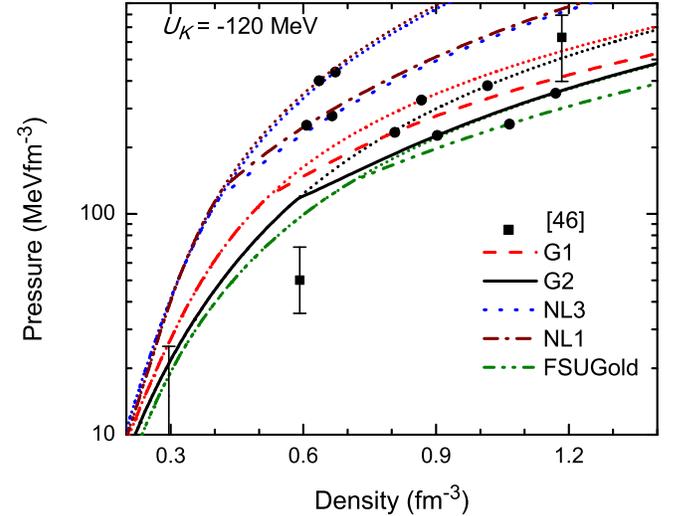}
\caption{(Color online) The EoS for non-kaonic and kaonic NS matter obtained
from various parameter
sets. The calculations done without considering kaons are represented
by the small-dotted lines whereas the calculations with kaons are represented
by different line patterns as given in the legends. The solid circles correspond to the values at
the center of maximum mass NS. Solid squares represent the
observational extraction \cite{expt_m5}, however not uniquely constrained
\cite{fattoyev.055803}.}\label{pd}
\end{figure}

\begin{figure}[t]
\includegraphics[width=.99\columnwidth]{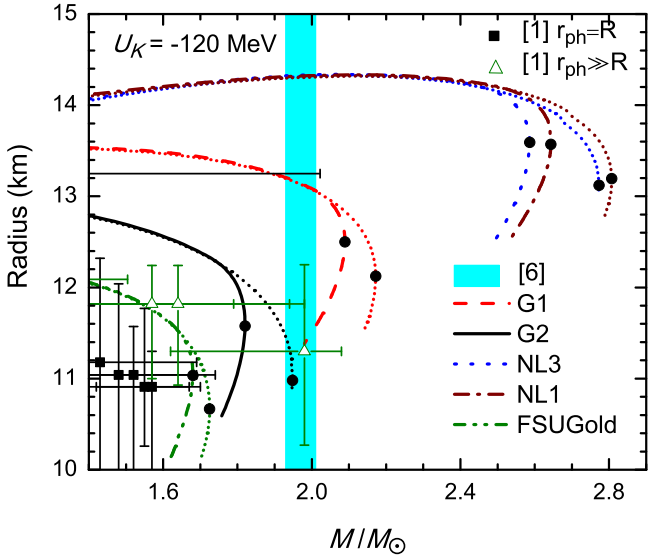}
\caption{(Color online) The mass-radius relation for non-kaonic and kaonic phases using different
parameter sets. The calculations done without considering kaons are represented
by the small-dotted lines.  The solid circles represent the maximum mass
in every case.  Mass is given in units of solar mass $M_\odot$. Solid squares
($r_{ph} = R$) and open triangles ($r_{ph}\gg R$) represent the
observational constraints \cite{Steiner33}, where $r_{ph}$ is the photospheric radius. The shaded region correspond to the recent observation of $1.97\pm0.04M_\odot$
star \cite{demorest}.}\label{mr}
\end{figure}

\begin{figure}[t]
\includegraphics[width=.99\columnwidth]{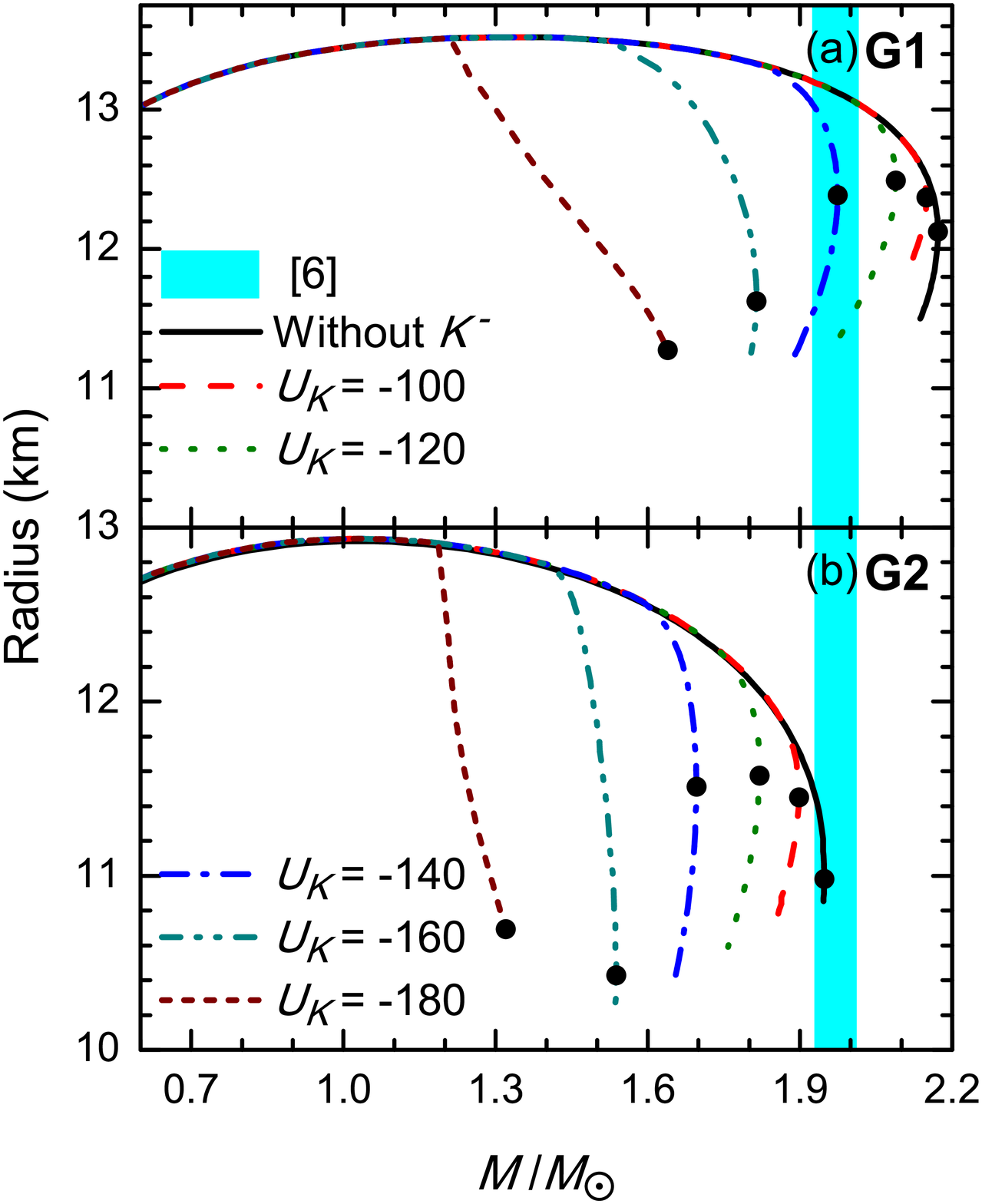}
\caption{(Color online) The mass-radius relation for NS calculated using different optical potentials and with G2
parameter set. Mass is  in units of solar mass $M_\odot$ and $U_K$ are
in MeV. The solid circles represent the maximum mass
in every case.}\label{mr_u}
\end{figure}

\begin{figure}[t]
\includegraphics[width=.99\columnwidth]{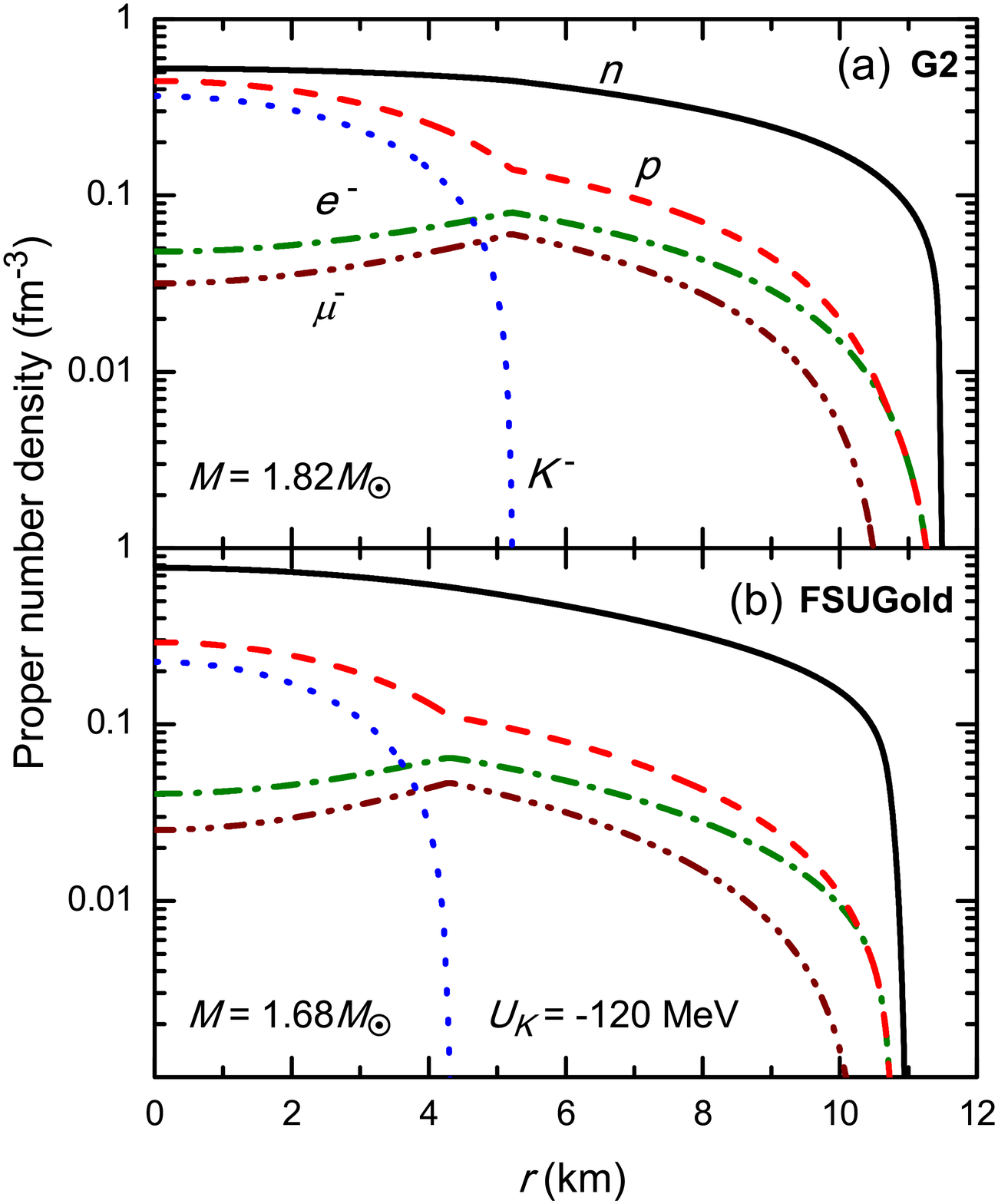}
\caption{(Color online) The composition of maximum mass NS as a function of radial distance
calculated
using  FSUGold and G2 parameter sets.}\label{dr}
\end{figure}

\subsection{Comparison between different interactions}

So far we have discussed our results with the G2 parameter set and in this
section we analyze how our results compare with different interactions considered
in this work. In Fig.~\ref{cp} we show the chemical potentials which indicate
the onset of $K^-$-condensation for different parameters. The maximum value
of $\mu_e\ (\mu_e^{max})$ denotes the point at which the transition
happens and is seen to be larger for G1 and G2. $\mu_e^{max}$  is almost
same for NL1 and FSUGold and the least value is for NL3.
$\mu_e^{max}$ also quantifies the number density of electrons which falls
sharply when $K^-$ starts to appear [See Fig.~\ref{densityg2}].
Both $\mu_e^{max}$ and the $\mu_n$  values at which the transition happens $(\mu_n^c)$, strongly
depend on the interaction. $\mu_n^c$ for G2 and FSUGold are more closer
resulting from a closer EoS. A sharper kink in Fig.~\ref{cp},
leading to a situation where there could be more than one value of $\mu_e$
for a given $\mu_n$, would have indicated a first order phase transition.
It is clear that for  $U_K=-120$ MeV, all the interactions lead only to a
second order phase transition where a mixed phase cannot appear.

In Fig.~\ref{density_g1} we compare the number densities arising from the
calculations with G1 and FSUGold, which can be compared with Fig.~\ref{densityg2} as well. The
results for G1 are very similar to that of G2 except for the early onset
of $K^-$-condensation in G1. Between Figs.~\ref{density_g1} (a) and (b),
we can see a significant difference in the relative population between protons
and neutrons at higher densities.   This is mainly
due to  the interplay between $\omega_K$ and $\mu_e$ which are plotted in
Fig.~\ref{kaon_para}. The decrease in $\omega_K$
or increase in $\mu_e$ favours more of $K^-$ and hence protons.  
$\omega_K$ varies linearly with the meson fields as given in Eq.~(\ref{eq:ke})
and should reflect the stiffness of the fields and hence that of the EoS. We can see in Fig.~\ref{kaon_para} that the $\omega_K$ for NL3 is quite different from others due to the fact that the EoS of NL3 is quite stiffer than others
\cite{ermf,fattoyev.055803}.  This stiffness arises from the vector potential
which grows almost as a straight line \cite{DelEstal:024314}. The quartic vector self-interaction brings
down the vector potential and makes the equation of state
and hence $\omega_K$ softer in other cases. The other quantity determining the onset of
$K^-$-condensation is $\mu_e (=\mu_n-\mu_p)$ which has terms similar to that
of the symmetry energy and is dominated by the $R_0$ field [Eq.~(\ref{eq:cp_np})]. Thus the density dependence of symmetry energy directly affects $\mu_e$ and
is crucial in determining the onset of $K^-$-condensation.  NL3, G1 and G2 yield similar symmetry energy (not shown here) and the unique isoscalar-isovector mixing ($\Lambda (g_\omega
V_0)^2(g_\rho R_0)^2$) in FSUGold suppresses the $R_0$ field which leads
to a softer symmetry energy and hence lesser $\mu_e$.

If we compare
between G2 and FSUGold, $\mu_e$(G2) $>\mu_e$(FSUGold) and with
similar $\omega_K$, kaons are more favoured for larger $\mu_e$.  Lesser kaons
in FSUGold leads to lesser number of protons at higher densities
in Fig.~\ref{density_g1}~(b). In Fig.~\ref{kaon_para} one can see that the
$\mu_e$ for NL3 and G1 are almost same till the transition point.  However
due to the large difference in $\omega_K$, which falls very sharply in the
case of NL3, $K^-$-condensation occurs at a smaller density in NL3 when compared
with G1.
These features can be seen from the EoS presented in Fig.~\ref{pd} as well.

Figure~\ref{pd} shows the pressure calculated with the different interactions,
with and without the inclusion of $K^-$. Similar to the case of symmetric
and pure neutron matter \cite{ermf,fattoyev.055803}, the higher order couplings
in G1, G2 and FSUGold lead to a softer EoS.  All the interactions suggest a
major change in pressure when we include $K^-$. The well known feature of
$K^-$ making the EoS softer, can also be seen clearly in Fig.~\ref{pd}. For
both kaonic and non-kaonic phases, the
difference between pressure obtained from G1 and G2 changes with density.
Around $\rho\sim0.5$ fm$^{-3}$, the difference is maximum and it decreases
as the density increase.  This is due to the interplay between the terms with
higher order couplings ($\kappa_3, \kappa_4, \eta_2$) which can give negative
contribution to pressure. Between G1 and G2,
the  sign of the coupling
constants $\kappa_4,\ \eta_2,$ and $\eta_\rho$  differ. Among these three couplings, the term comprising $\eta_\rho$ is weaker. In the case of G1,
due to the change in sign of $\kappa_4$, the combined effect of $\kappa_3$ and $\kappa_4$ on pressure is negligible.
Hence in G1, $\eta_2$ dominates at higher densities in reducing the pressure.
In case of G2, the contribution from terms having $\kappa_4$ and $\eta_2$
is negligible and $\kappa_3$ dominates at higher densities in reducing the
pressure. Thus in G1 and G2, $\eta_2$ (quartic scalar-vector cross-interaction), and $\kappa_3$ (cubic scalar
self-interaction) respectively, rule the suppression of pressure at very high densities.
$\eta_2$ being associated with the vector self-interaction is much stronger
at high densities and hence will be more effective than $\kappa_3$ in reducing
the pressure.  Couplings defined by $\eta_2$ and $\kappa_3$ are of different order and will naturally yield results of different curvature which leads
to the varying difference in pressure between G1 and G2. The difference in pressure from G2 and FSUGold is well reduced in the presence
of $K^-$, even at higher energies.  Interestingly, G2 with $K^-$ yields same
pressure as that of FSUGold without $K^-$ and in a broad energy range these results cease to differ. However, the corresponding energy densities are very
different (not shown here) and hence the resulting maximum mass NS have different central
densities (represented by solid circles in Fig.~\ref{pd}).  The EoS around the region of central
density is more dominant in determining the properties of NS. This region
is quite different for different parameters.  Also one can see that the central
density increases with softer EoS, and the presence of $K^-$ decreases
the central density and pressure.  

The change in central density ($\Delta\rho_c$) due to $K^-$    depends mostly on the following three 
quantities, viz., 
\begin{enumerate}
\item[(i)]
density at which $K^-$-condensation sets in ($\rho_B^K$) with 
$\Delta\rho_c \propto 1/\rho_B^K$,
\item[(ii)] the magnitude of central density with
$\Delta\rho_c \propto \rho_c$, and
\item[(iii)]
the stiffness of EoS at $\rho_c$, with stiffer EoS leading to larger $\Delta\rho_c$.
\end{enumerate}
The interplay of the above quantities gives arise to 
varying $\Delta\rho_c$ for different parameters.  For example, between G1 and G2,  quantities (ii) and (iii) dominate (i), resulting in a larger $\Delta\rho_c$
in case of G2. These effects should naturally be seen in the calculation of maximum mass and radius of NS.

We can obtain mass-radius relation for NS by solving
the well-known Tolman--Oppenheimer--Volkoff (TOV) equations \cite{tov,tov1}.
The results for mass-radius relation in NS are given in Fig.~\ref{mr} where
the calculations are done using  different interactions, without and with the $K^-$ ($U_K=-120$ MeV). The RMF models NL3 and NL1 suggest very large and massive NS. In these cases though the
maximum mass is reduced considerably (from $\sim$ 2.8 M$_\odot$ to $\sim$ 2.4 M$_\odot$) in the presence of kaons, these are overshooting the observational
constraints for NS \cite{Steiner33,expt_m,expt_m1,expt_m2,expt_m3,demorest}. Though G1 and G2 are from the same E-RMF model with same terms in the Lagrangian, their results for NS are quite different with G1 suggesting a larger and heavier NS. As explained earlier, G1 and G2 yield
EoS of different nature and have varying role of $K^-$. The result for G2 and FSUGold are the closest ones. Very interestingly, though their difference is larger when $K^-$ are not considered, the presence of $K^-$  brings these two results closer. This is the direct implication of the same feature we observed in
the case pressure (Fig.~\ref{pd}). The G2 interaction allows more $K^-$ to
be present (due to early onset of $K^-$-condensation) and hence makes the
EoS more softer which eventually get closer to that of FSUGold.         The
observed reduction of maximum mass due to kaons, in case of NL3 and FSUGold is consistent
with the values quoted in Ref.~\cite{sharma:23}. Overall, in the presence of $K^-$, the mass decreases and the radius corresponding
to maximum mass increases. All the observational constraints shown in Fig.~\ref{mr}
favor the models with higher order couplings but most of these constrains cover a broad range which
may not be precise enough to ascertain the presence of kaons.

In Fig.~\ref{mr_u}, we show the dependence of NS mass-radius on $U_K$ with
G2 parameter set. We
can see clearly that the stronger $U_K$ increases the contribution of $K^-$
which results in softer EoS which in turn yields a smaller and lighter NS.
The typical hook shape of the mass-radius curve is maintained till $U_K\lesssim-160
$ MeV and beyond that the curve takes a straight dip which is a characteristic
of first order $K^-$-condensation.  It is interesting to note that without
$K^-$, the maximum mass given by G2 ($1.95M_\odot$)\ is closer to the one suggested by Demorest \textit{et al.} ($1.97\pm0.04M_\odot$) \cite{demorest} with a radius of 11.03 km.  G1 with $U_K=-140$
MeV (graph not shown here) suggests a maximum mass of $1.97M_\odot$ with
a radius of 12.45 km. 
 
Figure~\ref{dr} shows the number density versus the radial distance from
the center of NS, calculated with G2 and FSUGold parameter sets, when the
NS is having the maximum mass ($1.68M_\odot$ for FSUGold, and $1.82M_\odot$ for G2 with $U_K=-120$ MeV). The behaviour of number densities shown in Fig.~\ref{dr}
is directly reflecting the patterns shown for relative populations plotted
against  density in Figs.~\ref{densityg2} and \ref{density_g1} (b). The difference between the radii of kaonic matter  suggested by G2 and FSUGold
is almost one km.  Another striking difference in the composition of NS
is the asymmetry of the core. G2 suggests a nearly symmetric matter at the
core whereas FSUGold indicates highly neutron-rich core.  We have seen that
this trend remains same even at higher values of  $U_K$ (not shown here).  It will be interesting to see how the presence of hyperons affect the overall scenario \cite{banik:055805}.  

\section{Summary \label{section:summary}}

In the present work, we have studied the  role of higher order couplings of extended RMF models on the onset and effect of kaon ($K^-$) condensation in neutron
stars (NS). We have calculated the NS properties with the successful RMF
parameters NL1 and NL3, the E-RMF parameters G1 and G2 (four additional couplings
to RMF), and the FSUGold parameters (two additional couplings
to RMF). 

In extended RMF models, with  most common values of kaon optical potential ($U_K\lesssim-160$
MeV), the transition from a non-kaonic phase to kaonic phase in NS has a character
of second order which rules out the possibility of having a mixed phase
of kaonic and non-kaonic matter. This is consistent
with the observation in Ref.~\cite{sanjay_kaon_t} where higher order
interactions are included using the lowest order chiral Lagrangian.
The role of more general higher order
operators was considered to be an open question and we have shown that they lead to a second order phase transition. This justifies the neglect of mixed phase
in our calculations mostly done with $U_K=-120$ MeV \cite{waas}.

Apart from the usual dependence
on $U_K$  reported elsewhere \cite{glen_kaon,sanjay_kaon_t},
the onset of kaon condensation in NS
strongly depends on the parameters of Lagrangian especially the higher order
couplings. This is due to the strong
variation in density dependence of the $K^-$ energy ($\omega_K$) and electron
chemical potential ($\mu_e$) whose interplay determine the onset of $K^-$-condensation.  Density dependence of $\omega_K$ is similar to that of EoS and $\mu_e$ varies in a way similar to symmetry energy.  So, any change in
the density dependence of EoS or that of symmetry energy will affect the
onset as well as the effect of $K^-$-condensation.
Without higher order couplings, NL1 and NL3 have stiffer EoS than others
which leads to a stiffer $\omega_K$ and hence a early onset of $K^-$-condensation.
 FSUGold comprises an unique isoscalar-isovector
mixing
which leads to a softer symmetry energy and hence a delayed onset of $K^-$-condensation.

 The central density of NS increases
with softer EoS, and the presence of $K^-$ decreases the
central density and pressure. The change in central density  due
to the presence of $K^-$ is different for different parameters due to the interplay between (i) the density defining onset of $K^-$-condensation, (ii) the central density ($\rho_c$) itself,
and (iii) the stiffness of EoS at $\rho_c$. Due to this, the impact of $K^-$
is more pronounced in G2 than in G1 and is weakest in FSUGold. All these effects are strongly
reflected in the calculation of mass-radius relation of NS.  The NS suggested
by models without higher order couplings (NL3 and NL1) contradict the observational constraints \cite{Steiner33,expt_m,expt_m1,expt_m2,expt_m3,demorest} even with the inclusion of \(K^{-}\). The higher order couplings play a dominant role (than kaons) in bringing the mass and radius of NS within observed
limits. Different parameter
sets lead to different concentration of kaons (which is known to affect the population
of protons \cite{glen_kaon}) in NS, and hence lead to different asymmetries in the core of NS. We have shown that G1 and G2 suggest a symmetric core whereas FSUGold suggests
a neutron rich core due to lesser amount of $K^-$ caused by an unique coupling which softens the symmetry energy. 

We conclude that the extended RMF models, which are quite successful in explaining several properties
of finite nuclei, suggest a strong influence of higher order couplings and
kaons in
NS whose properties are within the present observational constraints \cite{Steiner33,expt_m,expt_m1,expt_m2,expt_m3,demorest}.


\begin{thebibliography}{49}
\expandafter\ifx\csname natexlab\endcsname\relax\def\natexlab#1{#1}\fi
\expandafter\ifx\csname bibnamefont\endcsname\relax
  \def\bibnamefont#1{#1}\fi
\expandafter\ifx\csname bibfnamefont\endcsname\relax
  \def\bibfnamefont#1{#1}\fi
\expandafter\ifx\csname citenamefont\endcsname\relax
  \def\citenamefont#1{#1}\fi
\expandafter\ifx\csname url\endcsname\relax
  \def\url#1{\texttt{#1}}\fi
\expandafter\ifx\csname urlprefix\endcsname\relax\def\urlprefix{URL }\fi
\providecommand{\bibinfo}[2]{#2}
\providecommand{\eprint}[2][]{\url{#2}}

\bibitem[{\citenamefont{Steiner et~al.}(2010)\citenamefont{Steiner, Lattimer,
  and Brown}}]{Steiner33}
\bibinfo{author}{\bibfnamefont{A.~W.} \bibnamefont{Steiner}},
  \bibinfo{author}{\bibfnamefont{J.~M.} \bibnamefont{Lattimer}},
  \bibnamefont{and} \bibinfo{author}{\bibfnamefont{E.~F.} \bibnamefont{Brown}},
  \bibinfo{journal}{Astrophys. J} \textbf{\bibinfo{volume}{722}},
  \bibinfo{pages}{33} (\bibinfo{year}{2010}).

\bibitem[{\citenamefont{Champion et~al.}(2008)}]{expt_m}
\bibinfo{author}{\bibfnamefont{D.~J.} \bibnamefont{Champion}}
  \bibnamefont{et~al.}, \bibinfo{journal}{Science}
  \textbf{\bibinfo{volume}{320}}, \bibinfo{pages}{1309} (\bibinfo{year}{2008}).

\bibitem[{\citenamefont{Ransom et~al.}(2005)}]{expt_m1}
\bibinfo{author}{\bibfnamefont{S.~M.} \bibnamefont{Ransom}}
  \bibnamefont{et~al.}, \bibinfo{journal}{Science}
  \textbf{\bibinfo{volume}{307}}, \bibinfo{pages}{892} (\bibinfo{year}{2005}).

\bibitem[{\citenamefont{Freire et~al.}(2008)\citenamefont{Freire, Wolszczan,
  van~den Berg, and Hessels}}]{expt_m2}
\bibinfo{author}{\bibfnamefont{P.~C.~C.} \bibnamefont{Freire}},
  \bibinfo{author}{\bibfnamefont{A.}~\bibnamefont{Wolszczan}},
  \bibinfo{author}{\bibfnamefont{M.}~\bibnamefont{van~den Berg}},
  \bibnamefont{and} \bibinfo{author}{\bibfnamefont{J.~W.~T.}
  \bibnamefont{Hessels}}, \bibinfo{journal}{Astrophys. J.}
  \textbf{\bibinfo{volume}{679}}, \bibinfo{pages}{1433} (\bibinfo{year}{2008}).

\bibitem[{\citenamefont{van~der Meer et~al.}(2007)\citenamefont{van~der Meer,
  Kaper, van Kerkwijk, Heemskerk, and van~den Heuvel}}]{expt_m3}
\bibinfo{author}{\bibfnamefont{A.}~\bibnamefont{van~der Meer}},
  \bibinfo{author}{\bibfnamefont{L.}~\bibnamefont{Kaper}},
  \bibinfo{author}{\bibfnamefont{M.~H.} \bibnamefont{van Kerkwijk}},
  \bibinfo{author}{\bibfnamefont{M.~H.~M.} \bibnamefont{Heemskerk}},
  \bibnamefont{and} \bibinfo{author}{\bibfnamefont{E.~P.~J.}
  \bibnamefont{van~den Heuvel}}, \bibinfo{journal}{Astron. Astrophys.}
  \textbf{\bibinfo{volume}{473}}, \bibinfo{pages}{523} (\bibinfo{year}{2007}).

\bibitem[{\citenamefont{Demorest et~al.}(2010)\citenamefont{Demorest, Pennucci,
  Ransom, Roberts, and Hessels}}]{demorest}
\bibinfo{author}{\bibfnamefont{P.~B.} \bibnamefont{Demorest}},
  \bibinfo{author}{\bibfnamefont{T.}~\bibnamefont{Pennucci}},
  \bibinfo{author}{\bibfnamefont{S.~M.} \bibnamefont{Ransom}},
  \bibinfo{author}{\bibfnamefont{M.~S.~E.} \bibnamefont{Roberts}},
  \bibnamefont{and} \bibinfo{author}{\bibfnamefont{J.~W.~T.}
  \bibnamefont{Hessels}}, \bibinfo{journal}{Nature}
  \textbf{\bibinfo{volume}{467}}, \bibinfo{pages}{1081} (\bibinfo{year}{2010}).

\bibitem[{\citenamefont{Lattimer and Prakash}(2001)}]{lattimer:426}
\bibinfo{author}{\bibfnamefont{J.~M.} \bibnamefont{Lattimer}} \bibnamefont{and}
  \bibinfo{author}{\bibfnamefont{M.}~\bibnamefont{Prakash}},
  \bibinfo{journal}{Astrophys. J.} \textbf{\bibinfo{volume}{550}},
  \bibinfo{pages}{426} (\bibinfo{year}{2001}).

\bibitem[{\citenamefont{Friedman and Pandharipande}(1981)}]{nonrel}
\bibinfo{author}{\bibfnamefont{B.}~\bibnamefont{Friedman}} \bibnamefont{and}
  \bibinfo{author}{\bibfnamefont{V.~R.} \bibnamefont{Pandharipande}},
  \bibinfo{journal}{Nucl. Phys. A} \textbf{\bibinfo{volume}{361}},
  \bibinfo{pages}{502 } (\bibinfo{year}{1981}).

\bibitem[{\citenamefont{Wiringa et~al.}(1988)\citenamefont{Wiringa, Fiks, and
  Fabrocini}}]{nonrel1}
\bibinfo{author}{\bibfnamefont{R.~B.} \bibnamefont{Wiringa}},
  \bibinfo{author}{\bibfnamefont{V.}~\bibnamefont{Fiks}}, \bibnamefont{and}
  \bibinfo{author}{\bibfnamefont{A.}~\bibnamefont{Fabrocini}},
  \bibinfo{journal}{Phys. Rev. C} \textbf{\bibinfo{volume}{38}},
  \bibinfo{pages}{1010} (\bibinfo{year}{1988}).

\bibitem[{\citenamefont{Akmal and Pandharipande}(1997)}]{nonrel2}
\bibinfo{author}{\bibfnamefont{A.}~\bibnamefont{Akmal}} \bibnamefont{and}
  \bibinfo{author}{\bibfnamefont{V.~R.} \bibnamefont{Pandharipande}},
  \bibinfo{journal}{Phys. Rev. C} \textbf{\bibinfo{volume}{56}},
  \bibinfo{pages}{2261} (\bibinfo{year}{1997}).

\bibitem[{\citenamefont{M\"uller and Serot}(1996)}]{rel}
\bibinfo{author}{\bibfnamefont{H.}~\bibnamefont{M\"uller}} \bibnamefont{and}
  \bibinfo{author}{\bibfnamefont{B.~D.} \bibnamefont{Serot}},
  \bibinfo{journal}{Nucl. Phys. A} \textbf{\bibinfo{volume}{606}},
  \bibinfo{pages}{508} (\bibinfo{year}{1996}).

\bibitem[{\citenamefont{{M{\"u}ther} et~al.}(1987)\citenamefont{{M{\"u}ther},
  {Prakash}, and {Ainsworth}}}]{dbhf}
\bibinfo{author}{\bibfnamefont{H.}~\bibnamefont{{M{\"u}ther}}},
  \bibinfo{author}{\bibfnamefont{M.}~\bibnamefont{{Prakash}}},
  \bibnamefont{and} \bibinfo{author}{\bibfnamefont{T.~L.}
  \bibnamefont{{Ainsworth}}}, \bibinfo{journal}{Phys. Lett. B}
  \textbf{\bibinfo{volume}{199}}, \bibinfo{pages}{469} (\bibinfo{year}{1987}).

\bibitem[{\citenamefont{{Engvik} et~al.}(1996)\citenamefont{{Engvik}, {Osnes},
  {Hjorth-Jensen}, {Bao}, and {Ostgaard}}}]{dbhf1}
\bibinfo{author}{\bibfnamefont{L.}~\bibnamefont{{Engvik}}},
  \bibinfo{author}{\bibfnamefont{E.}~\bibnamefont{{Osnes}}},
  \bibinfo{author}{\bibfnamefont{M.}~\bibnamefont{{Hjorth-Jensen}}},
  \bibinfo{author}{\bibfnamefont{G.}~\bibnamefont{{Bao}}}, \bibnamefont{and}
  \bibinfo{author}{\bibfnamefont{E.}~\bibnamefont{{Ostgaard}}},
  \bibinfo{journal}{Astrophys. J.} \textbf{\bibinfo{volume}{469}},
  \bibinfo{pages}{794} (\bibinfo{year}{1996}).

\bibitem[{\citenamefont{{Glendenning} and {Moszkowski}}(1991)}]{rel_hyperon}
\bibinfo{author}{\bibfnamefont{N.~K.} \bibnamefont{{Glendenning}}}
  \bibnamefont{and} \bibinfo{author}{\bibfnamefont{S.~A.}
  \bibnamefont{{Moszkowski}}}, \bibinfo{journal}{Phys. Rev. Lett.}
  \textbf{\bibinfo{volume}{67}}, \bibinfo{pages}{2414} (\bibinfo{year}{1991}).

\bibitem[{\citenamefont{Pandharipande and Smith}(1975)}]{nonrel_pion}
\bibinfo{author}{\bibfnamefont{V.~R.} \bibnamefont{Pandharipande}}
  \bibnamefont{and} \bibinfo{author}{\bibfnamefont{R.~A.} \bibnamefont{Smith}},
  \bibinfo{journal}{Nucl. Phys. A} \textbf{\bibinfo{volume}{237}},
  \bibinfo{pages}{507 } (\bibinfo{year}{1975}).

\bibitem[{\citenamefont{Prakash et~al.}(1995)\citenamefont{Prakash, Cooke, and
  Lattimer}}]{rel_hy_quark}
\bibinfo{author}{\bibfnamefont{M.}~\bibnamefont{Prakash}},
  \bibinfo{author}{\bibfnamefont{J.~R.} \bibnamefont{Cooke}}, \bibnamefont{and}
  \bibinfo{author}{\bibfnamefont{J.~M.} \bibnamefont{Lattimer}},
  \bibinfo{journal}{Phys. Rev. D} \textbf{\bibinfo{volume}{52}},
  \bibinfo{pages}{661} (\bibinfo{year}{1995}).

\bibitem[{\citenamefont{Glendenning and Schaffner-Bielich}(1999)}]{glen_kaon}
\bibinfo{author}{\bibfnamefont{N.~K.} \bibnamefont{Glendenning}}
  \bibnamefont{and}
  \bibinfo{author}{\bibfnamefont{J.}~\bibnamefont{Schaffner-Bielich}},
  \bibinfo{journal}{Phys. Rev. C} \textbf{\bibinfo{volume}{60}},
  \bibinfo{pages}{025803} (\bibinfo{year}{1999}).

\bibitem[{\citenamefont{Glendenning and Schaffner-Bielich}(1998)}]{glen_prl}
\bibinfo{author}{\bibfnamefont{N.~K.} \bibnamefont{Glendenning}}
  \bibnamefont{and}
  \bibinfo{author}{\bibfnamefont{J.}~\bibnamefont{Schaffner-Bielich}},
  \bibinfo{journal}{Phys. Rev. Lett.} \textbf{\bibinfo{volume}{81}},
  \bibinfo{pages}{4564} (\bibinfo{year}{1998}).

\bibitem[{\citenamefont{Uchida et~al.}(2003)\citenamefont{Uchida, Sakaguchi,
  Itoh, Yosoi, Kawabata, Takeda, Yasuda, Murakami, Ishikawa, Taki
  et~al.}}]{Uchida:12}
\bibinfo{author}{\bibfnamefont{M.}~\bibnamefont{Uchida}},
  \bibinfo{author}{\bibfnamefont{H.}~\bibnamefont{Sakaguchi}},
  \bibinfo{author}{\bibfnamefont{M.}~\bibnamefont{Itoh}},
  \bibinfo{author}{\bibfnamefont{M.}~\bibnamefont{Yosoi}},
  \bibinfo{author}{\bibfnamefont{T.}~\bibnamefont{Kawabata}},
  \bibinfo{author}{\bibfnamefont{H.}~\bibnamefont{Takeda}},
  \bibinfo{author}{\bibfnamefont{Y.}~\bibnamefont{Yasuda}},
  \bibinfo{author}{\bibfnamefont{T.}~\bibnamefont{Murakami}},
  \bibinfo{author}{\bibfnamefont{T.}~\bibnamefont{Ishikawa}},
  \bibinfo{author}{\bibfnamefont{T.}~\bibnamefont{Taki}}, \bibnamefont{et~al.},
  \bibinfo{journal}{Phys. Lett. B} \textbf{\bibinfo{volume}{557}},
  \bibinfo{pages}{12 } (\bibinfo{year}{2003}).

\bibitem[{\citenamefont{Li et~al.}(2007)\citenamefont{Li, Garg, Liu, Marks,
  Nayak, Rao, Fujiwara, Hashimoto, Kawase, Nakanishi et~al.}}]{Li:162503}
\bibinfo{author}{\bibfnamefont{T.}~\bibnamefont{Li}},
  \bibinfo{author}{\bibfnamefont{U.}~\bibnamefont{Garg}},
  \bibinfo{author}{\bibfnamefont{Y.}~\bibnamefont{Liu}},
  \bibinfo{author}{\bibfnamefont{R.}~\bibnamefont{Marks}},
  \bibinfo{author}{\bibfnamefont{B.~K.} \bibnamefont{Nayak}},
  \bibinfo{author}{\bibfnamefont{P.~V.~M.} \bibnamefont{Rao}},
  \bibinfo{author}{\bibfnamefont{M.}~\bibnamefont{Fujiwara}},
  \bibinfo{author}{\bibfnamefont{H.}~\bibnamefont{Hashimoto}},
  \bibinfo{author}{\bibfnamefont{K.}~\bibnamefont{Kawase}},
  \bibinfo{author}{\bibfnamefont{K.}~\bibnamefont{Nakanishi}},
  \bibnamefont{et~al.}, \bibinfo{journal}{Phys. Rev. Lett.}
  \textbf{\bibinfo{volume}{99}}, \bibinfo{pages}{162503}
  (\bibinfo{year}{2007}).

\bibitem[{\citenamefont{Horowitz and
  Piekarewicz}(2001{\natexlab{a}})}]{horowitz:062802}
\bibinfo{author}{\bibfnamefont{C.~J.} \bibnamefont{Horowitz}} \bibnamefont{and}
  \bibinfo{author}{\bibfnamefont{J.}~\bibnamefont{Piekarewicz}},
  \bibinfo{journal}{Phys. Rev. C} \textbf{\bibinfo{volume}{64}},
  \bibinfo{pages}{062802} (\bibinfo{year}{2001}{\natexlab{a}}).

\bibitem[{\citenamefont{Todd-Rutel and Piekarewicz}(2005)}]{FSU_para}
\bibinfo{author}{\bibfnamefont{B.~G.} \bibnamefont{Todd-Rutel}}
  \bibnamefont{and}
  \bibinfo{author}{\bibfnamefont{J.}~\bibnamefont{Piekarewicz}},
  \bibinfo{journal}{Phys. Rev. Lett.} \textbf{\bibinfo{volume}{95}},
  \bibinfo{pages}{122501} (\bibinfo{year}{2005}).

\bibitem[{\citenamefont{Sharma and Pal}(2009)}]{sharma:23}
\bibinfo{author}{\bibfnamefont{B.~K.} \bibnamefont{Sharma}} \bibnamefont{and}
  \bibinfo{author}{\bibfnamefont{S.}~\bibnamefont{Pal}},
  \bibinfo{journal}{Phys. Lett. B.} \textbf{\bibinfo{volume}{682}},
  \bibinfo{pages}{23} (\bibinfo{year}{2009}).

\bibitem[{\citenamefont{Gambhir et~al.}(1990)\citenamefont{Gambhir, Ring, and
  Thimet}}]{Gambhir:132}
\bibinfo{author}{\bibfnamefont{Y.~K.} \bibnamefont{Gambhir}},
  \bibinfo{author}{\bibfnamefont{P.}~\bibnamefont{Ring}}, \bibnamefont{and}
  \bibinfo{author}{\bibfnamefont{A.}~\bibnamefont{Thimet}},
  \bibinfo{journal}{Ann. Phys. (N.Y.)} \textbf{\bibinfo{volume}{198}},
  \bibinfo{pages}{132 } (\bibinfo{year}{1990}).

\bibitem[{\citenamefont{Furnstahl et~al.}(1997)\citenamefont{Furnstahl, Serot,
  and Tang}}]{furnstahl}
\bibinfo{author}{\bibfnamefont{R.~J.} \bibnamefont{Furnstahl}},
  \bibinfo{author}{\bibfnamefont{B.~D.} \bibnamefont{Serot}}, \bibnamefont{and}
  \bibinfo{author}{\bibfnamefont{H.-B.} \bibnamefont{Tang}},
  \bibinfo{journal}{Nucl. Phys. A} \textbf{\bibinfo{volume}{615}},
  \bibinfo{pages}{441 } (\bibinfo{year}{1997}).

\bibitem[{\citenamefont{Arumugam et~al.}(2004)\citenamefont{Arumugam, Sharma,
  Sahu, Patra, Sil, Centelles, and Vi\~nas}}]{ermf}
\bibinfo{author}{\bibfnamefont{P.}~\bibnamefont{Arumugam}},
  \bibinfo{author}{\bibfnamefont{B.~K.} \bibnamefont{Sharma}},
  \bibinfo{author}{\bibfnamefont{P.~K.} \bibnamefont{Sahu}},
  \bibinfo{author}{\bibfnamefont{S.~K.} \bibnamefont{Patra}},
  \bibinfo{author}{\bibfnamefont{T.}~\bibnamefont{Sil}},
  \bibinfo{author}{\bibfnamefont{M.}~\bibnamefont{Centelles}},
  \bibnamefont{and} \bibinfo{author}{\bibfnamefont{X.}~\bibnamefont{Vi\~nas}},
  \bibinfo{journal}{Phys. Lett. B} \textbf{\bibinfo{volume}{601}},
  \bibinfo{pages}{51 } (\bibinfo{year}{2004}).

\bibitem[{\citenamefont{Kaplan and Nelson}(1986)}]{kap}
\bibinfo{author}{\bibfnamefont{D.~B.} \bibnamefont{Kaplan}} \bibnamefont{and}
  \bibinfo{author}{\bibfnamefont{A.~E.} \bibnamefont{Nelson}},
  \bibinfo{journal}{Phys. Lett. B} \textbf{\bibinfo{volume}{175}},
  \bibinfo{pages}{57 } (\bibinfo{year}{1986}).

\bibitem[{\citenamefont{Norsen and Reddy}(2001)}]{sanjay_kaon}
\bibinfo{author}{\bibfnamefont{T.}~\bibnamefont{Norsen}} \bibnamefont{and}
  \bibinfo{author}{\bibfnamefont{S.}~\bibnamefont{Reddy}},
  \bibinfo{journal}{Phys. Rev. C} \textbf{\bibinfo{volume}{63}},
  \bibinfo{pages}{065804} (\bibinfo{year}{2001}).

\bibitem[{\citenamefont{Pons et~al.}(2000)\citenamefont{Pons, Reddy, Ellis,
  Prakash, and Lattimer}}]{sanjay_kaon_t}
\bibinfo{author}{\bibfnamefont{J.~A.} \bibnamefont{Pons}},
  \bibinfo{author}{\bibfnamefont{S.}~\bibnamefont{Reddy}},
  \bibinfo{author}{\bibfnamefont{P.~J.} \bibnamefont{Ellis}},
  \bibinfo{author}{\bibfnamefont{M.}~\bibnamefont{Prakash}}, \bibnamefont{and}
  \bibinfo{author}{\bibfnamefont{J.~M.} \bibnamefont{Lattimer}},
  \bibinfo{journal}{Phys. Rev. C} \textbf{\bibinfo{volume}{62}},
  \bibinfo{pages}{035803} (\bibinfo{year}{2000}).

\bibitem[{\citenamefont{Wang et~al.}(2007)\citenamefont{Wang, Fu, and
  Liu}}]{kaon_wang}
\bibinfo{author}{\bibfnamefont{G.-h.} \bibnamefont{Wang}},
  \bibinfo{author}{\bibfnamefont{W.-j.} \bibnamefont{Fu}}, \bibnamefont{and}
  \bibinfo{author}{\bibfnamefont{Y.-x.} \bibnamefont{Liu}},
  \bibinfo{journal}{Phys. Rev. C} \textbf{\bibinfo{volume}{76}},
  \bibinfo{pages}{065802} (\bibinfo{year}{2007}).

\bibitem[{\citenamefont{Estal et~al.}(1999)\citenamefont{Estal, Centelles, and
  Vi\~nas}}]{DelEstal:443}
\bibinfo{author}{\bibfnamefont{M.~D.} \bibnamefont{Estal}},
  \bibinfo{author}{\bibfnamefont{M.}~\bibnamefont{Centelles}},
  \bibnamefont{and} \bibinfo{author}{\bibfnamefont{X.}~\bibnamefont{Vi\~nas}},
  \bibinfo{journal}{Nucl. Phys. A} \textbf{\bibinfo{volume}{650}},
  \bibinfo{pages}{443 } (\bibinfo{year}{1999}).

\bibitem[{\citenamefont{Estal et~al.}(2001{\natexlab{a}})\citenamefont{Estal,
  Centelles, Vi\~nas, and Patra}}]{DelEstal:024314}
\bibinfo{author}{\bibfnamefont{M.~D.} \bibnamefont{Estal}},
  \bibinfo{author}{\bibfnamefont{M.}~\bibnamefont{Centelles}},
  \bibinfo{author}{\bibfnamefont{X.}~\bibnamefont{Vi\~nas}}, \bibnamefont{and}
  \bibinfo{author}{\bibfnamefont{S.~K.} \bibnamefont{Patra}},
  \bibinfo{journal}{Phys. Rev. C} \textbf{\bibinfo{volume}{63}},
  \bibinfo{pages}{024314} (\bibinfo{year}{2001}{\natexlab{a}}).

\bibitem[{\citenamefont{Estal et~al.}(2001{\natexlab{b}})\citenamefont{Estal,
  Centelles, Vi\~nas, and Patra}}]{DelEstal:044321}
\bibinfo{author}{\bibfnamefont{M.~D.} \bibnamefont{Estal}},
  \bibinfo{author}{\bibfnamefont{M.}~\bibnamefont{Centelles}},
  \bibinfo{author}{\bibfnamefont{X.}~\bibnamefont{Vi\~nas}}, \bibnamefont{and}
  \bibinfo{author}{\bibfnamefont{S.~K.} \bibnamefont{Patra}},
  \bibinfo{journal}{Phys. Rev. C} \textbf{\bibinfo{volume}{63}},
  \bibinfo{pages}{044321} (\bibinfo{year}{2001}{\natexlab{b}}).

\bibitem[{\citenamefont{Sil et~al.}(2004)\citenamefont{Sil, Patra, Sharma,
  Centelles, and Vi\~nas}}]{Sil:044315}
\bibinfo{author}{\bibfnamefont{T.}~\bibnamefont{Sil}},
  \bibinfo{author}{\bibfnamefont{S.~K.} \bibnamefont{Patra}},
  \bibinfo{author}{\bibfnamefont{B.~K.} \bibnamefont{Sharma}},
  \bibinfo{author}{\bibfnamefont{M.}~\bibnamefont{Centelles}},
  \bibnamefont{and} \bibinfo{author}{\bibfnamefont{X.}~\bibnamefont{Vi\~nas}},
  \bibinfo{journal}{Phys. Rev. C} \textbf{\bibinfo{volume}{69}},
  \bibinfo{pages}{044315} (\bibinfo{year}{2004}).

\bibitem[{\citenamefont{Roca-Maza et~al.}(2008)\citenamefont{Roca-Maza,
  Centelles, Salvat, and Vi\~nas}}]{Roca-Maza:044332}
\bibinfo{author}{\bibfnamefont{X.}~\bibnamefont{Roca-Maza}},
  \bibinfo{author}{\bibfnamefont{M.}~\bibnamefont{Centelles}},
  \bibinfo{author}{\bibfnamefont{F.}~\bibnamefont{Salvat}}, \bibnamefont{and}
  \bibinfo{author}{\bibfnamefont{X.}~\bibnamefont{Vi\~nas}},
  \bibinfo{journal}{Phys. Rev. C} \textbf{\bibinfo{volume}{78}},
  \bibinfo{pages}{044332} (\bibinfo{year}{2008}).

\bibitem[{\citenamefont{Shukla et~al.}(2007)\citenamefont{Shukla, Sharma,
  Chandra, Arumugam, and Patra}}]{Shukla:034601}
\bibinfo{author}{\bibfnamefont{A.}~\bibnamefont{Shukla}},
  \bibinfo{author}{\bibfnamefont{B.~K.} \bibnamefont{Sharma}},
  \bibinfo{author}{\bibfnamefont{R.}~\bibnamefont{Chandra}},
  \bibinfo{author}{\bibfnamefont{P.}~\bibnamefont{Arumugam}}, \bibnamefont{and}
  \bibinfo{author}{\bibfnamefont{S.~K.} \bibnamefont{Patra}},
  \bibinfo{journal}{Phys. Rev. C} \textbf{\bibinfo{volume}{76}},
  \bibinfo{pages}{034601} (\bibinfo{year}{2007}).

\bibitem[{\citenamefont{Glendenning}(2007)}]{gle}
\bibinfo{author}{\bibfnamefont{N.~K.} \bibnamefont{Glendenning}},
  \emph{\bibinfo{title}{Compact Stars}} (\bibinfo{publisher}{Springer-Verlag,
  New York}, \bibinfo{year}{2007}), \bibinfo{edition}{2nd} ed.

\bibitem[{\citenamefont{Reinhard et~al.}(1986)\citenamefont{Reinhard, Rufa,
  Maruhn, Greiner, and Friedrich}}]{nl1_para}
\bibinfo{author}{\bibfnamefont{P.~G.} \bibnamefont{Reinhard}},
  \bibinfo{author}{\bibfnamefont{M.}~\bibnamefont{Rufa}},
  \bibinfo{author}{\bibfnamefont{J.}~\bibnamefont{Maruhn}},
  \bibinfo{author}{\bibfnamefont{W.}~\bibnamefont{Greiner}}, \bibnamefont{and}
  \bibinfo{author}{\bibfnamefont{J.}~\bibnamefont{Friedrich}},
  \bibinfo{journal}{Z. Phys. A} \textbf{\bibinfo{volume}{323}},
  \bibinfo{pages}{13} (\bibinfo{year}{1986}).

\bibitem[{\citenamefont{Lalazissis et~al.}(1997)\citenamefont{Lalazissis,
  K\"onig, and Ring}}]{nl3_para}
\bibinfo{author}{\bibfnamefont{G.~A.} \bibnamefont{Lalazissis}},
  \bibinfo{author}{\bibfnamefont{J.}~\bibnamefont{K\"onig}}, \bibnamefont{and}
  \bibinfo{author}{\bibfnamefont{P.}~\bibnamefont{Ring}},
  \bibinfo{journal}{Phys. Rev. C} \textbf{\bibinfo{volume}{55}},
  \bibinfo{pages}{540} (\bibinfo{year}{1997}).

\bibitem[{\citenamefont{Koch}(1994)}]{koch}
\bibinfo{author}{\bibfnamefont{V.}~\bibnamefont{Koch}}, \bibinfo{journal}{Phys.
  Lett. B} \textbf{\bibinfo{volume}{337}}, \bibinfo{pages}{7 }
  (\bibinfo{year}{1994}).

\bibitem[{\citenamefont{Waas and Weise}(1997)}]{waas}
\bibinfo{author}{\bibfnamefont{T.}~\bibnamefont{Waas}} \bibnamefont{and}
  \bibinfo{author}{\bibfnamefont{W.}~\bibnamefont{Weise}},
  \bibinfo{journal}{Nucl. Phys. A} \textbf{\bibinfo{volume}{625}},
  \bibinfo{pages}{287 } (\bibinfo{year}{1997}).

\bibitem[{\citenamefont{Fattoyev et~al.}(2010)\citenamefont{Fattoyev, Horowitz,
  Piekarewicz, and Shen}}]{fattoyev.055803}
\bibinfo{author}{\bibfnamefont{F.~J.} \bibnamefont{Fattoyev}},
  \bibinfo{author}{\bibfnamefont{C.~J.} \bibnamefont{Horowitz}},
  \bibinfo{author}{\bibfnamefont{J.}~\bibnamefont{Piekarewicz}},
  \bibnamefont{and} \bibinfo{author}{\bibfnamefont{G.}~\bibnamefont{Shen}},
  \bibinfo{journal}{Phys. Rev. C} \textbf{\bibinfo{volume}{82}},
  \bibinfo{pages}{055803} (\bibinfo{year}{2010}).

\bibitem[{\citenamefont{Danielewicz et~al.}(2002)\citenamefont{Danielewicz,
  Lacey, and Lynch}}]{expt_data}
\bibinfo{author}{\bibfnamefont{P.}~\bibnamefont{Danielewicz}},
  \bibinfo{author}{\bibfnamefont{R.}~\bibnamefont{Lacey}}, \bibnamefont{and}
  \bibinfo{author}{\bibfnamefont{W.~G.} \bibnamefont{Lynch}},
  \bibinfo{journal}{Science} \textbf{\bibinfo{volume}{298}},
  \bibinfo{pages}{1592} (\bibinfo{year}{2002}).

\bibitem[{\citenamefont{Horowitz and
  Piekarewicz}(2001{\natexlab{b}})}]{horowitz:5647}
\bibinfo{author}{\bibfnamefont{C.~J.} \bibnamefont{Horowitz}} \bibnamefont{and}
  \bibinfo{author}{\bibfnamefont{J.}~\bibnamefont{Piekarewicz}},
  \bibinfo{journal}{Phys. Rev. Lett.} \textbf{\bibinfo{volume}{86}},
  \bibinfo{pages}{5647} (\bibinfo{year}{2001}{\natexlab{b}}).

\bibitem[{\citenamefont{Gueorguiev et~al.}(2002)\citenamefont{Gueorguiev,
  Ormand, Johnson, and Draayer}}]{kappa_4}
\bibinfo{author}{\bibfnamefont{V.~G.} \bibnamefont{Gueorguiev}},
  \bibinfo{author}{\bibfnamefont{W.~E.} \bibnamefont{Ormand}},
  \bibinfo{author}{\bibfnamefont{C.~W.} \bibnamefont{Johnson}},
  \bibnamefont{and} \bibinfo{author}{\bibfnamefont{J.~P.}
  \bibnamefont{Draayer}}, \bibinfo{journal}{Phys. Rev. C}
  \textbf{\bibinfo{volume}{65}}, \bibinfo{pages}{024314}
  (\bibinfo{year}{2002}).

\bibitem[{\citenamefont{\"Ozel et~al.}(2010)\citenamefont{\"Ozel, Baym, and
  G\"uver}}]{expt_m5}
\bibinfo{author}{\bibfnamefont{F.}~\bibnamefont{\"Ozel}},
  \bibinfo{author}{\bibfnamefont{G.}~\bibnamefont{Baym}}, \bibnamefont{and}
  \bibinfo{author}{\bibfnamefont{T.}~\bibnamefont{G\"uver}},
  \bibinfo{journal}{Phys. Rev. D} \textbf{\bibinfo{volume}{82}},
  \bibinfo{pages}{101301} (\bibinfo{year}{2010}).

\bibitem[{\citenamefont{Oppenheimer and Volkoff}(1939)}]{tov}
\bibinfo{author}{\bibfnamefont{J.~R.} \bibnamefont{Oppenheimer}}
  \bibnamefont{and} \bibinfo{author}{\bibfnamefont{G.~M.}
  \bibnamefont{Volkoff}}, \bibinfo{journal}{Phys. Rev.}
  \textbf{\bibinfo{volume}{55}}, \bibinfo{pages}{374} (\bibinfo{year}{1939}).

\bibitem[{\citenamefont{Tolman}(1939)}]{tov1}
\bibinfo{author}{\bibfnamefont{R.~C.} \bibnamefont{Tolman}},
  \bibinfo{journal}{Phys. Rev.} \textbf{\bibinfo{volume}{55}},
  \bibinfo{pages}{364} (\bibinfo{year}{1939}).

\bibitem[{\citenamefont{Banik and Bandyopadhyay}(2001)}]{banik:055805}
\bibinfo{author}{\bibfnamefont{S.}~\bibnamefont{Banik}} \bibnamefont{and}
  \bibinfo{author}{\bibfnamefont{D.}~\bibnamefont{Bandyopadhyay}},
  \bibinfo{journal}{Phys. Rev. C.} \textbf{\bibinfo{volume}{64}},
  \bibinfo{pages}{055805} (\bibinfo{year}{2001}).

\end{thebibliography}
\end{document}